\documentclass[final,5p,times,twocolumn]{elsarticle}
\usepackage{graphicx}
\usepackage{amsmath}
\usepackage{amssymb}

\journal{Physica E}

\begin{document}
\begin{frontmatter}



\title{Nonequilirbium relations in spin glasses}


\author[label1]{Masayuki Ohzeki}
\author{Hidetoshi Nishimori$^b$}

\address[label1]{Department of Systems Science, Kyoto University, Yoshida-Honmachi, Sakyo-ku,
Kyoto 606-8501, Japan}
\address[label2]{Department of Physics, Tokyo Institute of Technology, Oh-okayama, Meguro-ku,
Tokyo 152-8551, Japan}
\begin{abstract}
The applications of nonequilbrium relations such as the Jarzynski equality and the fluctuation theorem to spin glasses are considered.
The spin glass is a basic platform where we consider an application of an approximate solver of combinatorial optimization problems, simulated annealing.
We find a novel relationship between an average through a nonequilibrium process where the temperature changes as in simulated annealing and a thermal average in equilibrium with different amounts of quenched randomness.
The results shown in the present study may serve as an alternative way to overcome critical slowing down in spin glasses.
It means that this way may mitigate difficulties in several hard optimization problems.
\end{abstract}

\begin{keyword}
Spin glasses \sep Jarzynski equality \sep Fluctuation theorem \sep Simulated annealing

\end{keyword}

\end{frontmatter}
\section{Introduction}
To reduce power loss in electric circuits, we have to minimize the circuit length.
This kind of problems are formulated into a more generic task to minimize or maximize a real single-valued function of multivariables. 
This is the optimization problem.
The cases in which variables take discrete values are known as combinatorial optimization problems \cite{OP}. 
To solve these problems is one of the most important tasks and has broad applications in science and engineering. 

One of the generic algorithms to tackle such problems is simulated annealing (SA), which has been developed to obtain the ground state of complex system such as the spin glasses, namely the minimization of the energy by stochastic searching in complicated phase space \cite{SA}. 
The system is driven by decreasing the temperature, which controls thermal fluctuations, according to a certain schedule. 
If the decrease speed is slow enough, the system stays at quasiequilibrium at each temperature.
We hope that the system reaches its ground state in the end.
However the existence of many local minima of the free energy of spin glasses leads to a plenitude of time scales which make it difficult to reach the equilibrium state, the global minimum, experimentally or in numerical simulations. 
It is thus sometimes difficult to find a correct solution in the protocol of SA due to the long equilibration time on spin glasses. 
Most of the optimization problems are expected to have similar difficulties since they can be mapped into searching problem of the ground state of spin glasses.

The above fact implies that the system during the process of SA applied to difficult problems is not close to their equilibrium. 
We therefore search for useful tools in nonequilibrium statistical physics to mitigate the difficulties.
During the last decade, a number of exact relations have been derived for nonequilibrium processes. 
The Jarzynski equality is one of these remarkable results. 
It shows that a statistical quantity depending on the work $W$ performed on a system in contact with a heat reservoir at an inverse temperature $\beta$ during a nonequilibrium process, $\exp \left( -\beta W\right) $, is related to the free energy difference $\Delta F$ between two equilibrium states, which are determined by the initial and final Hamiltonians of the system \cite{J1,J2}. 
This nonequilibrium relation holds for any intermediate schedule to change the parameter contained in the Hamiltonian. 
As one of the algorithms in numerical simulations, a practical use of the nonequilibrium relations has been proposed by several researchers \cite{NJ,Iba,Pop1}.
In the present work, we consider analytical applications of several nonequilibrium relations to spin glasses and establish the fundamental theories in order to overcome difficulties attributed to the critical slowing down in those systems.

We especially apply the nonequilibrium relations to spin glasses with the gauge symmetry for which we can exactly calculate several quantities \cite{HN81,HNbook}. 
As a result, we can obtain several exact relations between the nonequilibrium processes for spin glasses starting in a special subspace and the system properties in equilibrium characterized by different parameters from the properties at the starting point.

The present paper consists of five sections. In Section 2, the gauge symmetry of the spin-glass model is reviewed. 
In the next section, we introduce the nonequilibrium relations, the Jarzynski equality and a more generic theory, the Crooks fluctuation theorem \cite{J3,J4}. 
After then, we evaluate the nonequilibrium relations on the Nishimori line in Section 4. 
In the last section, we give a summary of the obtained results.
\section{Spin glasses}

\subsection{Model}
Let us consider the following simple spin-glass model, known as the random-bond Ising model, whose Hamiltonian is 
\begin{equation}
H=-\sum_{\langle ij\rangle }J_{ij}S_{i}(t)S_{j}(t),\label{H1}
\end{equation}%
where $S_{i}(t)$ is the Ising spin taking values $\pm 1$. 
The summation is taken over all bonds.
One may suppose usual nearest neighboring bonds on a $d$-dimensional hypercubic lattice. 
We make no restrictions on the type or the dimension of the lattice in the present paper. 
We can consider several types of bond distributions for the quenched randomness $J_{ij}$ such as the Gaussian distribution. 
We restrict ourselves to the case of $J_{ij}$ taking $\pm J$ for simplicity. 
The following analyses are straightforwardly applied to other cases. 

We separate the sign from the interaction $J_{ij}$ as $\tau _{ij}$, which takes $\pm 1$, below.
We combine the strength $J$ of the interactions with the inverse temperature $\beta$ as $K=\beta J$. 
The distribution functions for $\tau _{ij}$ can be written as \cite{HN81,HNbook} 
\begin{eqnarray}
P(\tau _{ij}) &=&p\delta (\tau _{ij}-1)+(1-p)\delta (\tau _{ij}+1)  \notag \\
&=&\frac{\mathrm{e}^{K_{p}\tau _{ij}}}{2\cosh K_{p}},\label{P2}
\end{eqnarray}%
where $\exp (-2K_{p})=(1-p)/p$. 
In the analysis of spin glasses, we need to evaluate not only the thermal average as $\left\langle \cdots \right\rangle_{K}$, but also the configurational average $[\cdots ]_{K_{p}}$ for $\tau_{ij}$ as
\begin{equation}
\left[ g(\{\tau _{ij}\})\right] _{K_{p}}=\sum_{\{\tau _{ij}\}}g(\{\tau
_{ij}\})\prod_{\langle ij\rangle }\frac{\mathrm{e}^{K_{p}\tau _{ij}}}{2\cosh
K_{p}}.
\end{equation}

\subsection{Gauge transformation for spin glasses}
Here we introduce the gauge transformation in spin glasses \cite{HN81,HNbook}. 
The gauge transformations for functions depending on $\mathrm{S}_{i}(t)$ and $\tau $ are defined as 
\begin{eqnarray}
S_{i}(t)&\rightarrow& S_{i}(t)\sigma _{i} \\
\tau_{ij}&\rightarrow& \tau_{ij}\sigma _{i}\sigma _{j}.
\end{eqnarray}%
For instance, the instantaneous Hamiltonian (\ref{H1}) is invariant under the gauge transformation, while the bond distribution function (\ref{P2}) is changed to 
\begin{equation}
P(\tau_{ij} )=\frac{\mathrm{e}^{K_p \tau_{ij}\sigma_i \sigma_j}}{2 \cosh K_p 
}.  \label{BF}
\end{equation}
Another important property is the invariance of summation over $\{S_i(t)\}$ and $\{\tau_{ij}\}$, since the gauge transformation merely changes the order of the summation.

Here we briefly give the general scheme of the analysis in random-bond Ising systems by the gauge transformation. 
If we take the configurational average of the gauge-invariant quantity $g(\{\tau_{ij}\})$, we find that 
\begin{equation}
\left[ g(\{\tau_{ij}\} )\right] _{K_{p}}=\sum_{\{\tau_{ij}\}
}g(\{\tau_{ij}\} )\prod_{\langle ij\rangle }\frac{\mathrm{e}^{K_p
\tau_{ij}\sigma_i \sigma_j}}{(2 \cosh K_p)^{N_B} },  \label{CA1}
\end{equation}
where $N_B$ expresses the number of bonds.
Since the left-hand side of this equation is independent of $\sigma $, the right-hand side is unchanged if we consider other configurations of $\sigma $. 
Therefore the summation over all possible configurations $\sigma $ yields the following relation with the coefficient consisting of the partition function with the coupling $K_{p}$ 
\begin{equation}
\left[ g(\{\tau_{ij}\} )\right] _{K_p}=\sum_{\{\tau_{ij}\} }g(\{\tau_{ij}\}
)\frac{Z(K_{p};\{\tau_{ij}\})}{2^{N}\left(2 \cosh K_p\right)^{N_B}},
\end{equation}
where $N$ denotes the number of sites. 
The partition function for the random bond Ising model is given as 
\begin{equation}
Z(K;\{\tau_{ij}\}) = \sum_{\sigma_i} \prod_{\langle ij \rangle}\mathrm{e}%
^{K\tau_{ij}\sigma_i \sigma_j}.
\end{equation}
Using the above properties of the gauge transformation, for instance, we can obtain the exact value of the internal energy on a special subspace $K=K_{p}$, known as the Nishimori line (NL), as \cite{HN81,HNbook}
\begin{equation}
\left[ \left\langle H\right\rangle _K\right] _{K}=-N_{B} J \tanh K.\label{ENL}
\end{equation}
As seen in this equation, the internal energy does not show any singularlities along NL.
We can evaluate an upper bound for the specific heat and some set of equalities and inequalities for the correlation functions on NL in similar ways.

\section{Nonequilibrium relations}
Let us consider the case that the system evolves following stochastic dynamics governed by the master equation. 
We change the value of the coupling $K$ from $K_0$ at $t=t_0$ to $K_f$ at $t=t_f$.
Correspondingly, the spin configuration $\{S_i(t)\}$ changes.
\subsection{Jarzynski equality}
The original version of the Jarzynski equality relates a quantity depending on the performed work during a nonequilibrium isothermal process and the free-energy difference between equilibrium states for the initial and final Hamiltonians \cite{J1,J2}. 
In the original formulation of JE, the work is defined as the energy difference due solely to the change of the Hamiltonian, and the heat is described by the energy difference through the change of the degrees of freedom in the system.
We can simply establish a variant version of JE in the case of changing the inverse temperature, which is useful in the following application, by defining the work as $-\beta W(\{S_i(t_k)\}) = - (K(t_{k+1})- K(t_k)) E(\{S_i(t_k)\})$, where $E(\{S_i(t)\})$ is the instantaneous energy for the spin configurations $\{S_i(t)\}$ given by the Hamiltonian (\ref{H1}) divided by $J$.
Then the Jarzynski equality can be written, similarly to the original version of JE, as
\begin{equation}
\left\langle e^{-\beta W}\right\rangle _{K_{0}\rightarrow K_{f}}=\mathrm{e}^{-\beta \Delta F}=\frac{Z(K_{f};\{\tau _{ij}\})}{Z(K_{0};\{\tau _{ij}\})}.
\label{JE1}
\end{equation}
The brackets $\langle \cdots \rangle _{A\rightarrow B}$ denote the average over various spin configurations realized in the nonequilibrium process starting from equilibrium at $t=t_0$.
We write the probability for the realized spin configurations from the initial spin configuration in equilibrium as $P_{K_{0}\rightarrow K_{f}}(\{S_{i}(t)\})$ by use of the solution of the master equation.
The average can then be expressed as 
\begin{eqnarray}\nonumber
& & \left\langle O(\{S_{i}(t)\})\right\rangle _{K_{0}\rightarrow K_{f}}\\
& & =\sum_{\{S_{i}(t)\}}P_{K_{0}\rightarrow
K_{f}}(\{S_{i}(t)\})O(\{S_{i}(t)\}),
\end{eqnarray}%
where $O(\{S_{i}(t)\})$ denotes an observable as a function of each realization of the spin configuration in a nonequilibrium process, and the summation extends over instantaneous spin configurations.
\subsection{Crooks fluctuation theorem}
The Crooks fluctuation theorem, which can be regarded as a more general version of the Jarzynski equality \cite{J3,J4}, relates $P_{K_{0}\rightarrow K_{f}}(\{S_i(t)\})$ with the probability $P_{K_{f} \rightarrow K_{0}}(\{S_i(t)\})$ of the inverse process as
\begin{equation}
\frac{P_{K_{0}\rightarrow K_{f}}(\{S_i(t)\})}{P_{K_{f}\rightarrow K_{0}}(\{S_i(t)\})} = \mathrm{e}^{\beta W- \beta \Delta F}.
\end{equation}
This relation yields \cite{Orland}
\begin{eqnarray}
& & \left\langle O(\{S_i(t)\})e^{-\beta W}\right\rangle _{K_{0}\rightarrow
K_{f}}  \notag \\
& & \quad =\langle O_{\rm r}(\{S_i(t)\})\rangle _{K_{f}\rightarrow K_{0}}\frac{%
Z(K_f;\{\tau_{ij}\} )}{Z(K_0;\{\tau_{ij}\})},  \label{JE2}
\end{eqnarray}
where $O_{\rm r}$ denotes the observable which depends on the backward process $K_{f}\rightarrow K_{0}$. 
A simple instance of $O(\{S_{i}(t)\})$ for the Ising spin system is the autocorrelation function $S_{i}(t_0)S_{i}(t_f)$. 
If we set an observable depending only on the final state, denoted by $O_{f}$, instead of $O(\{S_{i}(t)\})$, then $O_{\rm r}$ means an obsevable at the initial state in the backward process. 
Therefore $\langle O_{\rm r}\rangle_{K_{f}\rightarrow K_{0}}$ simply equals to the ordinary thermal average at the initial equilibrium state with the coupling constant $K_f$ as 
\begin{equation}
\left\langle O_{f}\mathrm{e}^{-\beta W}\right\rangle _{K_{0}\rightarrow
K_{f}}=\langle O\rangle _{K}\frac{Z(K_{f};\{\tau_{ij}\})}{%
Z(K_{0};\{\tau_{ij}\})}.  \label{JE3}
\end{equation}
These equations (\ref{JE1}), (\ref{JE2}) and (\ref{JE3}) can be proven under several formulations for a nonequilibrium process, the master equation, the Langevin equation, and the Fokker-Planck equation. 
We have employed the master-equation approach in Ref. \cite{J2} in the following analysis, since it is often used in the numerical simulation of spin glasses.
For the analysis of the master equation for Ising spin systems, several rules for the dynamics such as Metropolis \cite{Metro} and Glauber dynamics \cite{Glauber} are available, but we can prove that the following results are independent of the type of dynamics.
\section{Nonequilibrium relations on the Nishimori Line}
For a fixed bond configuration $\{\tau_{ij}\}$, we apply the nonequilibrium relation (\ref{JE3}) to gauge-invariant quantities denoted by $g(\{\tau_{ij}\})$. 
After then, we take the configurational average as follows, 
\begin{eqnarray}
& & \left[ \left\langle g_{f}(\{\tau_{ij}\})\mathrm{e}^{-\beta
W}\right\rangle_{K_0\rightarrow K_f}\right]_{K_{p}}  \notag \\
& & =\left[ \langle g(\{\tau_{ij}\})\rangle_{K_{f}}\frac{%
Z(K_{f};\{\tau_{ij}\})}{Z(K_{0};\{\tau_{ij}\})}\right] _{K_{p}}.  \label{GI1}
\end{eqnarray}
The quantity on the left-hand side means the configurational and nonequilibrium-process average of the observation of $g$ at the final time $t_f$, that is after the protocol $K_{0}\rightarrow K_{f}$. 
On the other hand, the quantity on the right-hand side means the configurational and thermal average of the equilibrium state for the final Hamiltonian. 
We use the gauge transformation to cancel the partition function of the denominator on the right-hand side in the above relation. 
The right-hand side of Eq. (\ref{GI1}) is rewritten explicitly as 
\begin{eqnarray}
&&\left[ \langle g(\{\tau_{ij}\})\rangle _{K_{f}}\frac{Z(K_{f};\{\tau_{ij}\})%
}{Z(K_{0};\{\tau_{ij}\})}\right]_{K_{p}}  \notag \\
&& =\sum_{\{\tau_{ij}\} }\frac{Z(K_{f};\{\tau_{ij}\})}{%
Z(K_{0};\{\tau_{ij}\})}\frac{\langle g(\{\tau_{ij}\})\rangle
_{K_{f}}\prod_{\langle ij\rangle }\mathrm{e}^{K_p \tau_{ij}\sigma_i \sigma_j}%
}{(2 \cosh K_p)^{N_B} }.  \notag \\
\end{eqnarray}%
All the quantities in this equation are invariant for the gauge transformation. 
We thus obtain
\begin{eqnarray}  \label{JENL1}
&&\left[ \langle g(\{\tau_{ij}\})\rangle _{K_{f}}\frac{Z(K_{f};\{\tau_{ij}\})%
}{Z(K_{0};\{\tau_{ij}\})}\right]_{K_{p}}  \notag \\
&& =\sum_{\{\tau_{ij}\} }\frac{\langle g(\{\tau_{ij}\})\rangle
_{K_{f}}Z(K_{p};\{\tau_{ij}\})}{2^N(2 \cosh K_p)^{N_B} }\frac{
Z(K_{f};\{\tau_{ij}\})}{Z(K_{0};\{\tau_{ij}\})}.  \notag \\
\end{eqnarray}
We here introduce the quantity $\left[ \langle g(\{\tau_{ij}\})\rangle_{K_{f}}\right]_{K_{p}}$. 
Similarly to the above calculation, we obtain the following identity by the gauge transformation to this quantity, 
\begin{eqnarray}
&&\left[ \langle g(\{\tau_{ij}\})\rangle _{K_{f}}\right] _{K_{p}}  \notag \\
&& =\sum_{\{\tau_{ij}\} }\frac{\langle g(\{\tau_{ij}\})\rangle
_{K_{f}}Z(K_{p};\{\tau_{ij}\})}{2^N(2 \cosh K_p)^{N_B} }.  \label{JENL2}
\end{eqnarray}%
Setting $K_{p}=K_{0}$ in Eq. (\ref{JENL1}) and $K_{p} = K_{f}$ in Eq. (\ref{JENL2}), we find a nonequilibrium relation for spin glasses by a simple algebra, 
\begin{eqnarray}
& & \left[ \langle g_f(\{\tau_{ij}\})\mathrm{e}^{-\beta W}\rangle _{K_0
\rightarrow K_{f}}\right]_{K_{0}}  \notag \\
& & =\left[ \langle g(\{\tau_{ij}\})\rangle _{K_{f}}\right]_{K_{f}}\left(%
\frac{2\cosh K_f}{2 \cosh K_0}\right)^{N_B}.  \label{JENL4}
\end{eqnarray}
If we set $g_{f}(\{\tau_{ij}\})=1$, we obtain the Jarzynski equality for spin glasses, 
\begin{equation}
\left[ \left\langle \mathrm{e}^{-\beta W}\right\rangle_{K_{0}\rightarrow
K_{f}}\right] _{K_{0}}=\left(\frac{2\cosh K_f}{2 \cosh K_0}\right)^{N_B}.
\label{JENL5}
\end{equation}
The relations (\ref{JENL4}) and (\ref{JENL5}) do not include a nontrivial free-energy difference as in ordinary non-equilibrium relations (\ref{JE1}), (\ref{JE2}), and (\ref{JE3}).
By Jensen's inequality, we derive the lower bound of the nonequilibrium work in terms of $\beta W$ on spin glasses with the gauge symmetry as
\begin{equation}
\left[ \beta \langle W \rangle_{K_{0} \to K_{f}}\right] _{K_{0}} \ge -N_B \log \left( \frac{2\cosh K_f}{2\cosh K_0}\right).
\end{equation}
All of the above results hold for any dimensional systems.
For spin glasses with the gauge symmetry, the lower bound of the nonequilibrium work is trivially given by the initial and final parameters.
This lower bound means the value for the case of the quasistatic process.
Even in the thermodynamic limit, in which the phase transition occurs, it does not possess any singularity.
We remark that the process does not go along NL.

If we substitute $g_f(\{\tau_{ij}\}) = H$ into Eq. (\ref{JENL4}) as an instance, we obtain 
\begin{eqnarray}
&&\left[ \left\langle H\mathrm{e}^{-\beta W}\right\rangle _{K_{0}\rightarrow
K_{f}}\right] _{K_{0}}  \notag \\
&& \quad = \left[ \langle H \rangle _{K_{f}}\right] _{K_{f}}\left(\frac{%
2\cosh K_f}{2 \cosh K_0}\right)^{N_B}.  \label{Ene}
\end{eqnarray}%
This equation shows that the internal energy after the cooling or heating process from a temperature on NL is proportional to the internal energy in the equilibrium state on NL corresponding to the final temperature.

\begin{figure}[tb]
\begin{center}
\includegraphics[width=50mm]{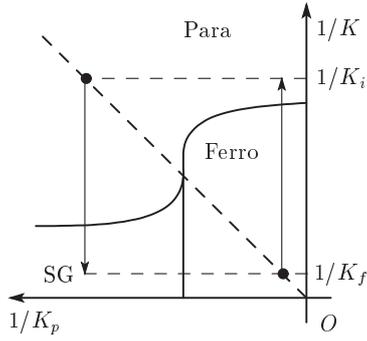}
\end{center}
\caption{{\protect\small Two related processes by the nonequilibrium relation (\ref{NE1}). 
Three phases (F: Ferromagnetic, P: paramagnetic, and SG: Spin Glass) are separated by a solid curve and a vertical line. 
The dashed and inclined line expresses NL.}}
\label{PDNL1}
\end{figure}

It is straightforward to obtain the nonequilibrium relation for the case of gauge-invariant quantities depending on the instantaneous spin configurations, for instance, the autocorrelation function $S_{i}(t_0)S_{i}(t_f)$ as, 
\begin{eqnarray}
&&\left[ \left\langle S_{i}(t_0)S_{i}(t_f )\mathrm{e}^{-\beta W}\right\rangle
_{K_{0}\rightarrow K_{f}}\right] _{K_{0}}  \notag \\
&&=\left[ \langle S_{i}(t_0)S_{i}(t_f)\rangle _{K_{f}\rightarrow K_{0}}%
\right] _{K_{f}}\left( \frac{2\cosh K_{f}}{2\cosh K_{0}}\right) ^{N_{B}}.\label{NE1}
\end{eqnarray}
This gives quite a mysterious relationship between the cooling and heating processes but with different amounts of quenched randomness characterized by $K_{0}$ and $K_{f}$. 
Let us consider the cooling process from a temperature on NL given by $(K_{0},K_{0})$ to a point away from NL $(1/K_{0},1/K_{f})$ as depicted by the down-pointing arrow in Fig. \ref{PDNL1}. 
Then the above equation relates this process with an inverse process from a temperature on NL $(1/K_{f},1/K_{f})$ to a point away from NL $(1/K_{f},1/K_{0})$ represented by the upward-pointing arrow in Fig. \ref{PDNL1}. 
The upward-pointing arrow process passes through the ferromagnetic and paramagnetic phases, whereas the cooling process goes through the spin glass phase with long-time equilibration. 
Its physical meaning should be studied in the future. 
This result may become an example to overcome the problem concerning long-time equilibration on spin glasses by the nonequilibrium relations. 

%-------------------------------------

\section{Summary}
We considered the application of the nonequilibrium relations to spin glasses with the gauge symmetry. 
The gauge symmetry enables us to change the non-trivial factor, the free energy difference, appearing in the nonequilibrium relations into a nontrivial factor given by the normalization factor of the quenched randomness. 

This study can be regarded as a practical application of the nonequilibrium relations to SA for spin glasses.
As one of the algorithms in numerical simulations, such a practical use of the nonequilibrium relations has been proposed by several researchers \cite{NJ,Iba,Pop1}.
We can implement our results in sush numerical methods and exploit them to investigate the equilibrium and nonequilibrium properties of spin glasses.
In that sense, our study may serve as a basis for the application of the nonequilibrium relations to spin glasses in numerical simulations.

Several nonequilibrium relations have been obtained by use of the gauge transformation for the dynamical system of spin glasses \cite{Ozeki1,Ozeki2,Ozeki3}.
However we emphasize that our results are established by involving the property of the Jarzynski equality and the fluctuation therem, differently from these papers, such as the independence of the schedule of the process and the relationship between two inverse processes.
In that sense, we hope that our results would open a way to mitigate the difficulties to search the ground state for spin glasses.

In this article, we showed the nonequilibrium relations only for the gauge-invariant quantities on NL. 
Other mysterious relations for non gauge-invariant quantities are reported elsewhere \cite{ON}.
\section*{Acknowledgment}
This work was financially supported by the 21st Century Global COE Program at Tokyo Institute of Technology `Nanoscience and Quantum Physics', and by CREST, JST.


\begin{thebibliography}{00}
\bibitem{OP} A. K. Hartmann and M. Weigt, Phase Transitions in
Combinatorial Optimization Problems: Basics, Algorithms and Statistical
Mechanics Wiley-VCH, Weinheim, (2005).

\bibitem{SA} S. Kirkpatrick, C. D. Gelett, and M. P. Vecchi, Science 220
(1983) 671.

\bibitem{J1} C. Jarzynski, Phys. Rev. Lett. \textbf{78}, 2690 (1997).

\bibitem{J2} C. Jarzynski, Phys. Rev. E \textbf{56}, 5018 (1997).

\bibitem{NJ} R. M. Neal, Statistics and Computing, \textbf{11} 125 (2001).

\bibitem{Iba} Y. Iba, Trans. Jpn. Soc. Artif. Intel.\textbf{16} 279 (2001).

\bibitem{Pop1} K. Hukushima, and Y. Iba, AIP. Conf. Proc. \textbf{690}, 200
(2003).

\bibitem{HN81} H. Nishimori, Prog. Theor. Phys. \textbf{66}, 1169 (1981).

\bibitem{HNbook} H. Nishimori, \emph{Statistical Physics of Spin Glasses and
Information Processing: An Introduction} (Oxford Univ. Press, Oxford, 2001).

\bibitem{J3} G. E. Crooks, J. Stat. Phys. \textbf{90}, 1481 (1998); Phys.
Rev. E \textbf{60}, 2721 (1999).

\bibitem{J4} G. E. Crooks, Phys. Rev. E \textbf{61}, 2361 (2000).

\bibitem{Orland} K. Malick, M. Moshe, and H. Orland, e-print arXiv:0711.2059v2
\bibitem{Metro} N. Metropolis, A. W. Rosenbluth, and M. N. Rosenbluth, A. H.
Teller, and E. Teller, J. Chem. Phys. \textbf{21}, 1087 (1953).
\bibitem{Glauber} R. J. Glauber, J. Math. Phys. \textbf{4}, 294 (1963).
\bibitem{Ozeki1}Y. Ozeki, J. Phys. A: Math. Gen. {\bf 28} 3645 (1995).
\bibitem{Ozeki2}Y. Ozeki, J. Phys.: Condens. Matter {\bf 9} 11171 (1997).
\bibitem{Ozeki3}Y. Ozeki, J. Phys. A; Math. Gen. {\bf 36} 2673 (2003). 
\bibitem{ON} M. Ohzeki, and H. Nishimori, e-print arXiv:1004.2389.
\end{thebibliography}
\end{document}